\documentclass[aip,jcp,graphicx,floatfix,showkey,twocolumn]{revtex4}

\usepackage{amsmath,bbm,graphicx,bm, multirow}

\begin{document}

\title{Simulation of tethered oligomers in nanochannels using
multi-particle collision dynamics}

\author{Riyad Chetram Raghu}
\email{rraghu@chem.utoronto.ca}

\author{Jeremy Schofield}
\email{jmschofi@chem.utoronto.ca}

\affiliation{Chemical Physics Theory Group,
Department of Chemistry,
University of Toronto, 80 Saint George Street, Ontario,
Canada, M5S 3H6}


\date{\today}

\begin{abstract}
The effect of a high Reynold's number, pressure-driven flow of a compressible gas on the conformation of an 
oligomer tethered to the wall of a square-channel is studied under both ideal 
solvent and poor solvent conditions using a hybrid multiparticle collision 
dynamics and molecular dynamics algorithm.  Unlike previous studies, the flow field contains an elongational component in addition to a shear component as well as fluid slip near the walls and results in a Schmidt number for the polymer beads that is less than unity.  In both solvent regimes the 
oligomer is found to extend in the direction of flow.  Under the ideal solvent conditions, torsional twisting of the chain and aperiodic cyclical dynamics are observed for the end of the oligomer.  Under poor solvent 
conditions, a metastable helix forms in the end of the chain despite the lack of 
any attractive potential between beads in the oligomeric chain.   The formation 
of the helix is postulated to be the result of a solvent induced chain collapse that has been confined to a single dimension by a strong flow field.
\end{abstract}

\maketitle

\section{ Introduction}

The effect of flow on the conformation of surface tethered polymers is an area of study of increasing pertinence  as advances in materials science, nanofabrication techniques and chemistry are incorporated into the engines of industry.    Whether tethered polymers are used to prevent protein fouling in biomedical devices (such as through PEGylation of blood wetting devices),  anchor receptors in lab-on-a-chip enzyme-linked immunosorbent assay (ELISA) \cite{Kim1996}, replace frictional contacts with viscous contacts through weeping lubrication \cite{Klein2006} or serve a structural purpose within the device (e.g. nylon submicron filters), understanding how a flow field modifies the behaviour of a polymer chain can improve the design and functionality of these devices \cite{Bagassi1989}.    While many of these devices consist of polymer layers, where intermolecular polymer interactions are common, isolated tethered polymer chains, where intermolecular polymer interactions are absent, can also be present in these devices.    

The behaviour of isolated tethered polymer chains under pure shear flow was examined theoretically well before an experimental apparatus could be devised to study the phenomenon.   The earliest model consisted of a Brownian dynamics simulation of a single bead connected by a Hookean spring to surface in a linear shear field \cite{DiMarzio1978}.  Beyond the intuitive prediction that the polymer would extend into the direction of flow and adopt a stretched configuration, this simple model predicted that instead of assuming a steady conformation in the flow field, the tethered bead would undergo cyclical dynamics that could be described as a four part cycle of \textit{entrainment}, \textit{rotation}, \textit{contraction} and \textit{extension}.  During the \textit{entrainment} step the bead is entrained in the flow field and the polymer chain is stretched.  During the \textit{rotation} step, the force on the bead is balanced by the force constant of the spring.  In the case of a polymer chain the effective "spring" is created by the entropic loss of the polymer chain as it is extended and loses configurational freedom.  As the bead is now fixed at a set distance from the adhesion point, it is rotated towards the wall of the channel.  In the \textit{contraction} step, as the bead is rotated towards the wall of the channel, it rotates out of the flow field and, no longer being entrained in the flow field, it retracts back towards the wall.  Finally, in the \textit{extension} step, as the bead nears the anchor point, the Brownian diffusion of the bead carries it away from the wall where it becomes entrained in the flow field again.  These dynamics were subsequently confirmed experimentally by studies of tethered DNA oligomers in aqueous solution \cite{LeDuc1999,Doyle2000} although the resolution of the data was insufficient to perform a quantitative analysis of the dynamics.

Cyclical dynamics also have been predicted by other theoretical models \cite{Zhang2009,Delgado2006,Dubbeldam2006,Holzer2006}.  When these dynamics are manifest, there is some question as to whether or not they are periodic in nature.  In a Brownian dynamics simulation of a FENE chain in a linear shear field the dynamics appeared to be periodic with a period roughly equal to $10\tau_R$, where $\tau_R$ is the relaxation time of the polymer chain \cite{Delgado2006}.  Using a simplified system of three beads bound by Hookean springs to the points of an equilateral triangle, the study of Holzer et al \cite{Holzer2006} found a Hopf bifurcation in the position of the beads as the shear rate was increased.  At low shear rates the polymer extended into the direction of flow, at intermediate shear rates a limit cycle emerged corresponding to periodic oscillations, which became aperiodic at higher shear rates.  In contrast, cyclical dynamics were not found in studies where a rigid bond model was used instead of a spring model \cite{Parnas1991} suggesting that bond elasticity may be necessary for this phenomenon to emerge.  More recently the study of Zhang et al \cite{Zhang2009} studied cyclical dynamics using three methods: Brownian Dyanamics, Lattice Boltzmann and mixed direct simulation Monte Carlo (DSMC)-Molecular Dynamics (MD) method.  They found that the dynamics were aperiodic and suggested that the cyclical dynamics may be caused by a Poisson process like the tumbling of untethered polymers under shear flow \cite{Das2008}.  If the tumbling process is a Poisson process, then the delay between tumbling events should follow an exponential distribution.  In addition to the Brownian dynamics approach, the effects of flow on tethered polymer chains have also been examined using Monte Carlo simulations \cite{Rutledge2001,Karaiskos2009,Lai1993}, MD  \cite{Dimitrov2007,Pierlioni1995,Edberg1986}, DSMC \cite{Peikos1996} and Multiparticle Collision Dynamics (MPCD) \cite{Malevanets2000a,Webster2005}.

The previously mentioned studies of the effect of a flow field on a tethered polymer only considered a pure shear flow.  However, there is often slip flow within these devices and there can be an elongational component to the flow if the fluid is compressible.  The effect of the compressibility of the medium is particularly important in gas flows.  As the contraction step of the cycle assumes that the fluid has no velocity near the wall, the presence of a slip velocity alone alters the dynamics.  

In this paper, the behavior of a tethered oligomer is examined in a pressure-driven flow field of a compressible gas under two different solvent conditions; a solvent condition where the oligomer assumes an equilibrium conformation in solvent that is equivalent to its vacuum configuration (ideal solvent), and a more compact configuration (poor solvent).  Under ideal solvent conditions it is shown that the cyclical dynamics are restricted to the end of the chain and are aperiodic under pressure-driven flow.   Under poor solvent conditions the oligomer adopts a helical configuration instead of undergoing cyclical dynamics.

To simulate the flow on the oligomer, the dynamics were studied using a hybrid MPCD\cite{Malevanets2000,Kapral2008} and MD algorithm.  A MD algorithm was necessary as the oligomer model made use of an explicit potential.  The MPCD algorithm was chosen as the interparticle distance was significant relative to the length scale over which dynamical variables changed and, as a particle based algorithm, it was easy to interface with a molecular dynamics algorithm.  The flow profile was modeled on a pressure driven compressible fluid flow within a nanofluidic channel using slip boundary conditions.  

The following section of the article, section 2, presents the oligomer model that was simulated in this article, the MPCD algorithm that was used to simulate the fluid and the dimensionless numbers that characterize the physical significance of these models.   The results for the ideal solvent conditions and poor solvent conditions are presented in sections 3 and 4 respectively.  The final section of the article, section 5, contains a summary of the results and the potential implications for practical applications.

\section{The Simulation Model}

The essential difficulty with the simulation of the effect of channel flow on the conformation of an oligomer is the existence of two disparate time and length scales within the system associated with evolution of the flow profile within the channel and the configuration of the oligomer.  Due to the elongational component of the flow field, a significant portion of the channel needs to be simulated to capture the flow profile. In particular, the entire cross-section of the channel must be simulated, as well as a sufficient portion of the channel both upstream and downstream of the oligomer to establish the flow field and accommodate inlet and outlet artefacts.  The flow variables such as pressure, density, temperature and velocity also require approximately a nanosecond to attain a steady state. This requires a continuum approach or a mesoscopic model.  However, to simulate the effect of this field on the oligomer, the oligomer must be represented by a model of sufficient complexity to capture subtle changes in the native conformation, which requires integration steps in the femtosecond range.  Thus the oligomer requires a MD integration approach.  As a result of these requirements, a hybrid MD-MPCD algorithm\cite{Malevanets2000} was used to approach this simulation.  The MPCD algorithm has the advantage of impressive speed of simulation while preserving hydrodynamics\cite{Malevanets1999} and, being particle based, it is straightforward to adapt the model to incorporate the MD subsystem for the oligomeric chains.  The oligomer model itself is simplified from a complete molecular description of a homopolymer.  It maintains the backbone structure of the chain but removes the side groups of the chain, incorporating them into the backbone beads.  This facilitates the use of a larger integration time step and decreases the number of variables that could affect the configuration of the oligomer itself, leading to results that have general applicability.  

\subsection{The MPCD-MD Algorithm}

Channel flow for a compressible fluid within a nanoscopic channel is complicated by several factors; the fluid is expanding due to an isentropic expansion as the pressure decreases, the fluid is being sheared by frictional interactions with the walls of the channel, and the fluid is slipping at the walls of the channels.  Due to the complexity of this flow, the flow profile was simulated using a MPCD algorithm, and a MD subsystem was introduced to model the interactions between the fluid and the oligomer.  

In order to streamline the standard molecular dynamics equations, the MPCD algorithm forgoes the use of an explicit potential between solvent particles.  The domain is subdivided into cells of edge length $L=1$ nm.  At discrete time intervals, $\tau_c$, particles sharing the same cell have their velocity rotated about the average cell velocity, $\overline{\mathbf{v}}$, by an angle $\phi$.  I.e. velocity of the $i^{th}$ particle in the cell, $\mathbf{v}^*_i$ is
\begin{equation*}
\mathbf{v}^*_i=\mathbf{R}_k\left(\pm\phi\right)
\left(\mathbf{v}_i-\overline{\mathbf{v}}\right)+
\overline{\mathbf{v}},
\end{equation*}
where $\mathbf{v}_i$ is the pre-rotation velocity of the particle and $\mathbf{R}_k\left(\pm\phi\right)$ is the matrix for a rotation through an angle $\phi$, about an axis $k$.   The magnitude of the rotation is kept constant during the simulation, for the simulations presented here $\phi=\pi/4$ .  In three dimensions there are two models for the choice of the direction about which the rotation is performed.  Either the rotation is performed about any randomly chosen direction \cite{Malevanets1999}, or the rotation is performed about a randomly chosen Cartesian axis of the system \cite{Ihle2001}. The forms of the transport coefficients are somewhat different between the two methods \cite{Tuzel2003,Ihle2004,Kikuchi2003,Hecht2005}.  The first model has the advantage of assuring the proper rotational symmetry for the stress tensor of the fluid, while the second method is slightly more efficient.  For the purposes of modeling channel flow the second model is sufficient, thus it is used here. For this fluid the transport coefficients are \cite{Tuzel2003,Ihle2004,Kikuchi2003,Hecht2005,Raghu2010}:
\begin{eqnarray*}
\mu & = & \frac{\rho k_BT\tau_c\left(3\rho_c^2+2\rho_c+1+\sqrt{2}\left(\rho_c^2+2^(-0.5)-2\right)
\right)}{2\left(3\rho_c^2-2\rho_c-1-\sqrt{2}\left(\rho_c^2+2^(-0.5)-2\right)
\right)}\\
&&+\frac{m\left(\rho_c-1+e^{-\rho_c}\right)\left(1-2^{-0.5}\right)}
{18\tau_c\L},\\
D_T &= & \frac{k_BT\tau_c\left(9\rho_c^3+\alpha_t\right)}{2m\left(9\rho_c^3-\alpha_t
\right)}+\frac{L^2\left(\rho_c-1\right)\left(1-2^{-0.5}\right)}
{15\tau_c\rho_c^2}\\
D &= &\frac{k_BT\tau_c}{2m}\left(\frac{3\rho_c}{\left(\rho_c-1+e^{-\rho_c}\right)
\left(1-2^{-0.5}\right)}-1\right),
\end{eqnarray*} 
where $m$ is the mass of the fluid particle, $\mu$ is the shear viscosity of the fluid, $D_T$ is the thermal diffusivity, $D$ is the self-diffusion coefficient of fluid particles, $\rho_c$ is the average number of particles per cell and 
\begin{eqnarray*}
\alpha_t &=&\left(2\left(\rho_c-1\right)\left(2^{-0.5}-1\right)-\rho_c\right)
\left(\rho_c^2+8^{0.5}\left(\rho_c-1\right)\right) \\
&&+9.6\left(2^{-0.5}-1\right)
\left(\rho_c-1\right).
\end{eqnarray*}
The algorithm also randomly shifts cells prior to rotation and shifts them back following a rotation by a random displacement vector $\mathbf{\delta}$  to eliminate the effect of momentum correlations from particles being in the same cells during subsequent rotations \cite{Ihle2001, Tuzel2003}, however as the mean free path length is greater than $L$, this procedure is not strictly required.

Although there is no local structure or excluded volume interactions in the MPCD model fluid, there is still momentum transfer between particles.  It is possible to introduce some excluded volume interactions by modifying the algorithm to introduce specular collisions between particles in cells where the average cell velocity indicates that the particles are travelling towards each other \cite{Ihle2006}.  A closed MPCD system still conserves mass, momentum and energy.  It can be shown through an H-theorem that an MPCD system evolves to a state of maximum entropy and, in the low rarefaction limit, reproduces hydrodynamics consistent with the Navier-Stokes equation \cite{Malevanets1999}.

For the purposes of introducing a physical connection between the simulation data and a real fluid, $\tau_c$ was selected to be equal to the mean collision time for the gas being modeled.  I.e.,
\begin{equation*}
\tau_c=\frac{2}{\pi r^2\left(P_{in}+P_{out}\right)}\sqrt{\frac{mk_BT}{3}},
\end{equation*}
 where $r=0.98$ $\AA$  is the particle radius $m=39.943$ AMU is the particle mass, $T=298.15$ K and $P_{in}$ and $P_{out}$ are the inlet and outlet pressure. $P_{out}$ was held constant for all simulations at $82.3$ MPa (corresponding to a particle density of $20$ particles/cell), while $P_{in}$ was varied to generate a pressure gradient. The MPCD algorithm used here was not integrated using the rotational period $\tau_c$. Instead a time step, $\Delta t=0.5$ $ps$ that was shorter than $\tau_c$ was used for numerical integration.  As  $\tau_c\overline{\mathbf{v}}>L$, a shorter time step was required for the stable application of the inlet and outlet boundary conditions.  The boundary conditions operated by modifying the particle distributions on a cell-by-cell basis at each time step, thus if the distribution changed significantly between intervals it would produce artefacts in the flow profile.  The use of a shorter time step also allowed the velocity to be truncated at $L/\Delta t$ to prevent particles from traveling more than one cell length in a single time step.   As $\Delta t<\frac{L}{4}\sqrt{\frac{m}{k_BT}}$, very few particles require truncation and having a screening range of particle motion simplified the construction of the MD susbsystem during each $\Delta t$ interval.  In the simulations of different pressure differences presented here, $\tau_c$ ranged from $8 \Delta t$ for equilibrium simulations to $4 \Delta t$ for the highest flow rate simulations. Measurements that involve the variance of the distribution were corrected for the omission of these few high velocity particles by multiplying the measured variance by the ratio of the variance of the complete Maxwell-Boltzmann distribution to the variance of the truncated Maxwell-Boltzmann distribution.

A leap-frog integration scheme with a time step $\delta t=\Delta t/1000$, was used to simulate the oligomer and its interactions with the fluid\cite{Frenkel2002}.  As the fluid particles do not interact directly with each other, the algorithm requires the integration of the equations of motion of only particles that could interact with the oligomer between $t$ and $t+\Delta t$. Due to the velocity truncation, the maximum distance a particle can be from the oligomer and still interact is $2L+r_c$, where $r_c=L$ is the cut off radius of the fluid-oligomer potential.  Hence, in order to interface the oligomer model with the MPCD channel, a cell envelope that contains all the particles that are within at least $2L+r_c$ of the oligomer is separated from the system and integrated with time step $\delta t$, creating a hybrid MD-MPCD system.  The overall technique is similar to that employed in Malevanets and Kapral \cite{Malevanets2000}, but simplified due to the velocity truncation.

Due to the complexities introduced by the elongational component of the flow field, mechanisms for generating a pure shear flow field such as sliding brick \cite{Evans1984} boundary conditions or a systemic force field \cite{Allahyarov2002} are not appropriate.  Instead, the flow field is created by a source-sink boundary condition algorithm that establishes a pressure gradient at the inlet and outlet via the addition and removal of particles from the inlet and outlet of the system \cite{Raghu2010}.  Particles were added from a Maxwell-Boltzmann distribution at a specific temperature to supply thermal control. The mean velocity, used for the calculation of the thermal velocity, was calculated using a series solution to the Navier-Stokes solution for the isothermal flow of an ideal gas in a square channel at the inlet, while a spatial and temporal local average was used in the outlet.  The interactions between the solvent particles and the wall were modelled by an adiabatic collision operator, where the component of particle velocity that is normal to the wall is reflected and the remaining components are rotated by an angle that is uniformly drawn from the $[0,2\pi)$ interval.  The algorithm is presented in full detail elsewhere\cite{Raghu2010}, complete with details on the boundary conditions and implementation. 

\begin{figure}
\includegraphics[width=\columnwidth]{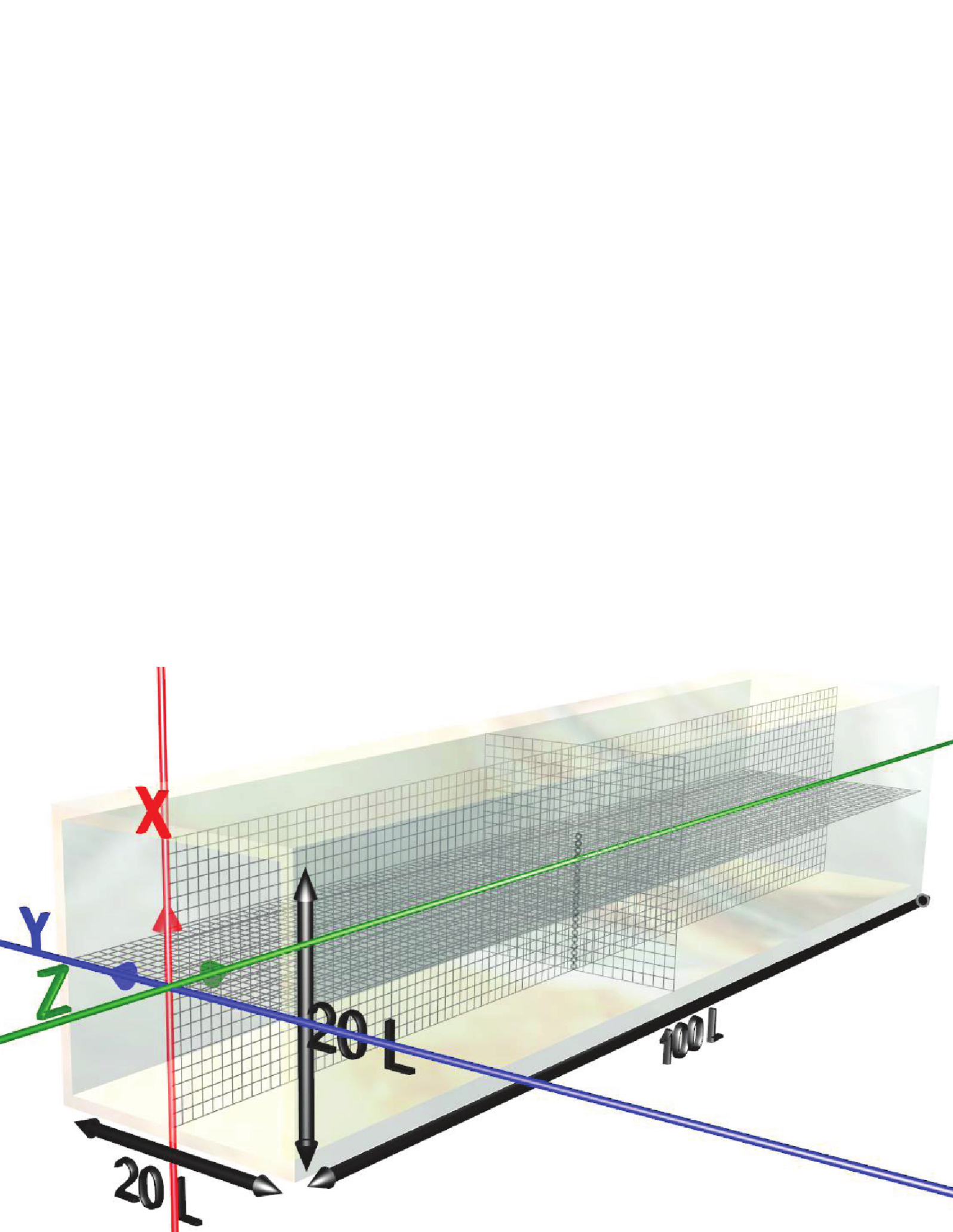}
\caption{Channel Geometry.  For these simulations H=20 L and L=100 L.  The oligomer is anchored at (-10 L, 0 L, 50 L).}
\label{fig:Fig1}
\end{figure}

The simulations were performed in a $20 L \times 20 L \times 100 L$ channel of square geometry (Fig.~\ref{fig:Fig1}) with the oligomer chain anchored to the centre of the channel at $(-10 L, 0 L, 50 L)$, with the $z$-axis representing the direction of flow.  Initially the oligomer is absent from the channel, which is initialized to a pressure corresponding to the inlet pressure.  The system is allowed to equilibrate for $2000$ $\Delta t$, corresponding to $1$ $ns$, and the flow profile is generated from a washout distribution, where the channel is initially filled to the higher inlet pressure and the gradient is allowed to form from the drainage of particles through the outlet.  This has the advantage of more rapid equilibration than the inverse fill-up distribution, where the system is equilibrated to the outlet pressure and particles are added through the inlet to create the gradient, and avoids the creation of shocks within the system.  At $2000$ $\Delta t$, the oligomer is inserted into the channel, with a straight configuration extended along the x axis, where each bead is separated from the previous by the equilibrium bond length.  This is done to remove some variation between runs and to observe the relaxation time of the oligomer.  Each bead is given an initial velocity drawn from a Maxwell-Boltzmann distribution with $\sqrt{k_BT/m}=1.2456\times 10^{-1}$ $L/\Delta t$.  Any fluid particles that have a non-zero potential with the oligomer chain at the time of insertion are deleted. 

\subsection{The Oligomer Model}

A critical factor determining the behavior of the oligomeric chain under shear flow is the structure and potentials used to model the oligomer.  As the MPCD solvent used in these simulations lacks short range order, a similarly simple oligomeric chain was chosen, where each repeat unit of the oligomer is replaced by single bead and each bead is connected to the subsequent bead through a spring potential.  This oligomeric chain is representative of an oligomer such as polyethylene or poly(tetrafluoroethlyene) where there are no large side groups on the chain and the repeat unit for the chain can be reduced to each backbone carbon (e.g. $CH_2$, $CF_2$).  Whenever physical data were required, the data for polyethylene were used to model the oligomer.  Additional potentials are defined between oligomer beads and: other oligomer beads that do not share a bond, MPCD fluid particles and the channel wall.  For these simulations each oligomer bead had a mass that was $0.3512$ $m$ and a radius of $0.050$ $L$.  The length of the oligomer was held constant at $100$ beads.  The parameters for the potentials described below are given in table~\ref{tab:Tab1}. 

The oligomer chain bond potential was modeled as a finitely extendable non-linear elastic (FENE) spring \cite{Warner1972} with a modification to allow for an equilibrium bond length. I.e. the bond potential was:
\begin{equation*}
u^{FENE}_{i,i+1}=K_{FENE} \left \vert a-\vert \mathbf{r}_{i,i+1}\vert
\frac{a}{2}\ln\left(\frac{\vert\mathbf{r}_{i,i+1}\vert}
{2a-\vert\mathbf{r}_{i,i+1}\vert}\right) \right \vert,
\end{equation*}
where $\mathbf{r}_{i,i+1}=\mathbf{r}_{i+1}-\mathbf{r}_{i}$ is the separation vector between the position of bead $i$ and $i+1$, $a$ is the equilibrium bond length (thus $0$ and $2a$ are the extensibility limits) and $K_{FENE}$ is the spring constant for the bond.  The potential is zero at the equilibrium bond length and infinite at a separation of either $0$ or $2a$.  Hence there is a limit on the extension of the oligomer, just as there would be for a physical bond.  In theory this limit is $2a$, though, due to the choice of parameter values, the bond length stays within the range of $0.9a$  to $1.1a$ under the specified simulation conditions.  This potential was also used to tether the oligomer to the wall at the anchor point.

The interactions between the oligomer beads and the fluid particles were modeled by a scaled Lennard-Jones 6-12 potential of the form:
\begin{equation*}
u^{LJ}_{i,j}=E_{min}\left(
\left(\frac{r_{min}}{\vert\mathbf{r}_{i,j}\vert}\right)^{12}
-2\left(\frac{r_{min}}{\vert\mathbf{r}_{i,j}\vert}\right)^{6}\right),
\end{equation*}
where $E_{min}$ is the energy at the minimum of the potential well and $r_{min}$ is the particle separation of minimum energy.  While the Lennard-Jones potential is a standard model for an inter-molecular potential, the attractive portion of the potential when used as an oligomer-solvent potential could act to cross-link the oligomer when there are no explicit solvent-solvent potentials present.  Solvent-solvent interactions prevent multiple solvent particles from occupying the same volume within interaction range of the oligomer.  If the attractive potential is strong enough to bind a solvent particle, it can bind multiple solvent particles to the same site and these solvent particles can act to cross-link the oligomeric chain.  Instead of a model that is closer to the physical reality of the situation, the model creates a magnified primary solvation shell with no secondary structure.  Within the parameter range chosen to model this interaction, this phenomena did not occur as there was no solvent density increase in the vicinity of the oligomer.

The Lennard-Jones potential was used to simulate both poor solvent and ideal solvent conditions. A ideal solvent is one in which the enthalpy of mixing for the solvent and the polymer is zero, and the polymer adopts the same conformation that it would in the absence of a solvent.  In theory an ideal solvent for a polymer would be a $\Theta$ solvent. Theoretically, under $\Theta$ solvent conditions the radius of gyration of a tethered polymer is expected to have a similar magnitude to an untethered polymer and scale with the length as $R_{gyr}=a\sqrt{N/6}$, where $N$ is the number of Kuhn segments in the chain and $a$ is the length of the segment\cite{Aksimentiev1999}.  However, unlike a true polymer, the excluded volume interactions in the core of a short-chained oligomer prevent the oligomer from adopting Gaussian chain characteristics \cite{Ladd1992} and the oligomer assumes a configuration at least as large as a self-avoiding chain\cite{Doi1986,Ladd1992}, $R_{gyr}\propto N^{0.59}$,  where the proportionality constant is specific to the oligomer being examined.  As such we use the term ideal solvent instead of $\Theta$ solvent and state that if the oligomer adopts a smaller conformation than its vacuum configuration, the solvent is said to be a poor solvent, while if it adopts a larger conformation in the solvent than in vacuum, the solvent is said to be a good solvent.       Hence the solvent conditions were estimated by comparing the radius of gyration of the oligomer in the solvent, to the radius of gyration of the oligomer in the absence of a solvent under equilibrium conditions, as opposed to comparing the size of the oligomer to Gaussian chain statistics.  The radius of gyration of an oligomer is written as, 
\begin{eqnarray*}
R_{gyr}&=&\sqrt{\frac{1}{N}\left(\left(\sum^N_{i=1}\mathbf{q}_i\ldotp\mathbf{q}_i
\right)-\left(\sum^N_{i=1}\mathbf{q}_i\right)^2\right)} \\
&&\sqrt{R^2_{gyr,x}+R^2_{gyr,y}+R^2_{gyr,z}},
\end{eqnarray*}
where $N$ is the number of beads in the oligomeric chain, $\mathbf{q}_i$ is the position vector of the $i^{th}$ bead in the polymer chain and $R_{gyr,x}$, $R_{gyr,y}$ and $R_{gyr,z}$ are the $x$, $y$ and $z$ components of the radius of gyration.  These components are analogous to the radius of gyration and are formed using the  $x$, $y$ or $z$ component of $\mathbf{q}_i$, i.e.
\begin{equation*}
R_{gyr,x}=\sqrt{\frac{1}{N}\left(\left(\sum^N_{i=1}q^2_{xi}\right)
-\left(\sum^N_{i=1}q_{xi}\right)^2\right)}.
\end{equation*}
In the absence of a solvent, a centimer was found to have a radius of gyration of $1.85 \pm 0.32$  $L$, while the radius of gyration was 
$1.79 \pm 0.26$ $L$ under ideal solvent conditions and $1.21 \pm 0.20$ $L$ under poor solvent conditions.

The interactions between beads that were not bonded within the oligomeric chain were determined by a truncated harmonic potential:
\begin{eqnarray*}
u^h_{i,j} =\left\{ 
\begin{array}{ll}
K_h r_c^2/2 \,  \left(1-{\vert \mathbf{r}_{ij}\vert}/
{r_c}\right)^2 & \vert\mathbf{r}_{ij}\vert<r_c \\
0 &\vert\mathbf{r}_{ij}\vert\geq r_c,
\end{array}
\right.
\end{eqnarray*}
where $K_h$ is the spring constant of the potential and $r_c$ is the cut off radius of the potential.  Without excluded volume interactions, an attractive potential tends to cause the chain to favor a collapsed configuration.  This constraint coupled with the need for a clean cut-off potential led to its selection.  

The interactions between the walls and the oligomer were modeled using an exponential repulsive potential:
\begin{eqnarray*}
u^e_{i,wall}=\alpha e^{-\beta d}-d\beta\alpha e^{-\beta d_c}&0<d\leq d_c,
\end{eqnarray*}
where $d$ is distance between the particle and the nearest point on the wall, $\alpha$ is the maximum potential between bead $i$ and the wall, $\beta$ is the decay constant of the potential and $d_c$ is the cut off radius of the potential.  This potential was chosen to prevent the oligomer from adsorbing to the wall of the channel.  

\begin{table}
\begin{tabular}{c|c}
\hline \rule{0pt}{20pt}
Parameter&Value\\
\hline \hline
$K_{FENE}$ & $43.7216$ $mL/\Delta t^2$ \\
$a$ & $0.154$ $L$ \\
$E_{min}$ (Ideal Solvent) & $-3.0483\times 10^{-3}$ $mL^2/\Delta t^2$  \\
$E_{min}$ (Poor Solvent) & $-1.0953\times 10^{-2}$ $mL^2/\Delta t^2$ \\
$r_{min}$ & $0.2073$ $L$ \\
$K_h$ & $25.0268$ $mL^2/\Delta t^2$ \\
$r_c$ & $0.308$ $L$ \\
$\alpha$ & $1.315\times 10^2$ $mL^2/\Delta t^2$ \\
$\beta$ & $0.3$ $L^{-1}$ \\
$d_c$ & $1$ $L$ \\
\hline
\end{tabular}
\caption{Potential parameter values for simulations.}
\label{tab:Tab1}
\end{table}

\subsection{Dimensionless Numbers}

Dimensionless numbers are often used in fluid mechanics and polymer physics to characterize the solvent and determine the dominant physical processes within the system for the purpose of modeling. Thus far this section has presented the simulation models and physical parameters, this section will discuss the physical significance of these choices.  

The fluid velocity profile is governed by four dimensionless numbers, the Knudsen number, $K_n$, the Reynolds number, $R_e$, the Mach number, $M_a$ and the Prandtl number $P_r$.  The first three of these numbers are related to each other through the equation,
\begin{equation*}
K_n=\sqrt{\frac{\pi C}{2}}\frac{M_a}{R_e},
\end{equation*}
where $C$ is the ratio of specific heats.  The Knudsen number is defined as $K_n=\lambda/\Lambda$ where $\Lambda=\rho\left(\frac{d\rho}{dz}\right)^{-1}$ is the characteristic length scale of the system\cite{Roy2003}.  In essence it indicates whether the inter-particle distance is significant over the range that hydrodynamic parameters vary significantly.  If $K_n<0.001$ then a continuum method such as the Navier-Stokes equation can be used to model the flow profile, while a large $K_n$ ($K_n>10$) indicates that a molecular dynamics approach is more appropriate to simulate the dynamics of the system.  In the simulations with flow presented here, the Knudsen number ranged from $\left(8\pm 3\right)\times 10^{-4}$ to $\left(2.08\pm 0.03\right)\times 10^{-2}$ in spatial
regions in the channel that are far from the polymer.  In the vicinity of
the polymer, local thermodynamic and transport properties are not
well-defined due to rapid changes in the microscopic definition of such quantities over molecular length scales.
The Mach number is the ratio of the speed of the fluid to the speed of sound in the fluid, and is defined for a gas as $M_a=v\sqrt{\frac{\rho m}{CP}}$.  A low Mach number indicates that compressibility effects of the fluid can be ignored.  For the present simulations, the Mach number ranged from $0$ to $0.2413\pm 0.0006$,
indicating that
the compressibility effects of the fluid cannot be universally ignored.  The Reynolds number is the ratio of inertial forces to viscous forces, and for a square channel it is defined as $R_e={m\rho \bar{v}H}/{\mu}$.  At high Reynolds numbers ($R_e>2300$ for pipe flow\cite{Welty1984}), turbulence begins to emerge in the flow.  Over the course of the simulations, the Reynolds number ranged from $0$ to $320\pm 1$, indicating laminar flow.  Finally, the Prandtl number represents the ratio of momentum diffusivity to thermal diffusivity, and is defined as $P_r= {\mu}/({m\rho D_T})$.  In the simulations analyzed here, the Prandtl number had a value of $4.4\pm 0.1$, indicating that convective mass transfer was dominant over conductive mass transfer. 

In the presence of a grafted polymer, another dimensionless number becomes relevant, the Schmidt number.  This number indicates the relative strength of convective mass transfer processes to diffusive mass transfer processes and also indicates whether hydrodynamic interactions are gas-like or liquid-like.  The Schmidt number, defined as $S_c={\mu}/{(\rho m D)}$, was computed to have a value of $0.313\pm 0.003$, indicating a gas-like solvent.  As the flow profile includes the pressure drop in the fluid and the SRD fluid described here has an ideal gas equation of state, a gas-like fluid should be simulated instead of a liquid-like fluid where the compressibility is negligible.   

The diffusion coefficient for the center of mass of the polymer, $D_p$, can be written as \cite{Mussawisade2005},
\begin{equation*} D_p=\frac{k_BT\tau_c}{M_bN}\left(\frac{1}{\eta}-\frac{1}{2}\right),
\end{equation*} 
where 
\begin{equation*}
\eta=\sum_{l=1}^\infty \frac{e^{-B_c}B_c^{l-1}}{\left(l-1\right)!}\frac{\rho_cm}{lm_b+\rho_cm}, 
\end{equation*}
$B_c$ is the average number of polymer beads per cell and $m_b$ is the mass of a polymer bead. The ratio of the diffusion coefficient of the fluid to the diffusion coefficient of the polymer , $D/D_p=34$ indicating that the center of mass of an unattached polymer diffuses significantly more slowly than the fluid corresponding to the
conditions relevant for Brownian dynamics of polymer as a whole.  Similarly, the ratio of the self-diffusion coefficients of the fluid and polymer bead particles,
$D/D_b$, can be estimated\cite{Ripoll2005} to be around $0.38$, which suggests that the polymer beads themselves move more rapidly than the surrounding fluid particles due to their small mass.  Under these solvent conditions, the dynamics of the polymer beads is not Brownian-like due to the small size and mass of the individual monomers in the oligomer.
     

\section{ideal solvent Conditions}

\begin{figure}
\includegraphics[width=0.9\columnwidth]{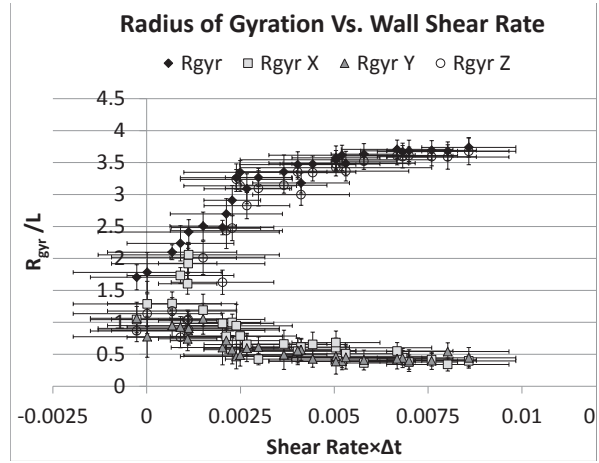}
\caption{$R_{gyr}$ and its three components vs. Wall Shear Rate at the point of adhesion (0, 10 L, 50 L).}
\label{fig:Fig2}
\end{figure}

The initial response of the oligomer from its initial stretched configuration to the fluid is dependent on two effects, an initial entropic retraction of the chain towards the anchor point as the oligomer attempts to assume a random coil configuration, and the drag force of the flow field pulling the chain in the direction of flow.  At low flow rates, the entropic retraction is dominant and the oligomer adopts a random coil configuration.  As the flow rate is increased, the oligomer initially assumes a parabolic arc that matches the velocity profile of the channel. Then, as the tension in the chain increases, the oligomer straightens and rotates until it lies recumbent near the wall of the channel. This manifests as a general increase in  $R_{gyr}$, more specifically an increase in the flow component, $R_{gyr,z}$, and a decrease in the other two components, $R_{gyr,x}$  and $R_{gyr,y}$.  Figure \ref{fig:Fig2} shows $R_{gyr}$ and its three components as a function of the shear rate at the adhesion point of the oligomer to the channel.  There appears to be a minimum velocity below which the oligomer does not extend in the direction of flow, and above this velocity the oligomer rapidly extends to a value that is approximately $80\%$ of its contour length.  Because there are no explicit attractive interactions between beads in the chain,  there is no impetus for adopting a specific conformation and the loss of conformational entropy from extension should be the primary resistance against oligomer elongation.  The entropic resistance of a material to physical deformation is the basis of the entropic theory of rubber elasticity.  Extrapolating this theory to a single chain, the relationship between the shear rate, $\dot{\gamma}$, and the elongation ratio of the oligomer, $\lambda$, should be \cite{Young1991},
\begin{equation*}
\dot{\gamma}\propto\left(\lambda-\frac{1}{\lambda^2}\right).
\end{equation*}
As the $\lambda^{-2}$ term vanishes very rapidly for large elongation ratios, one expects essentially a linear relationship between the elongation ratio and the shear rate.  However the shear force is not the only force acting on the oligomer, as the flow field has an additional contribution from the slip velocity.  At low flow rates, a greater than linear increase in the flow rate is expected.  At higher flow rates, the oligomer reaches its elastic limit as bond stretching forces become dominant.  Thus a sigmoidal relationship between the radius of gyration and the shear rate is expected and confirmed by the data in Fig.~\ref{fig:Fig2}. 

\begin{figure}[htb]
\includegraphics[width=\columnwidth]{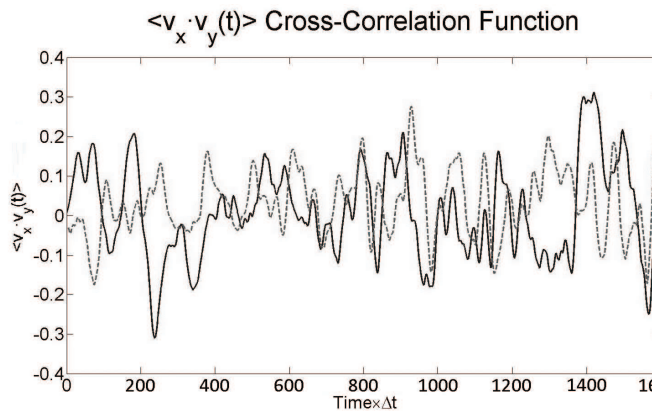}
\caption{Cross-correlation function between the $x$ and $y$ velocity components for the last $50$ beads of the centimer.  The solid line corresponds to a flow rate of $\left(3.26\pm 0.10\right)\times 10^4$  particles/$\Delta$t, which corresponds to a wall shear rate of $\left(2.3\pm 0.8\right)\times 10^{-3}$ $\Delta t^{-1}$ and slip velocity of $\left(2.9\pm 0.8\right)\times 10^{-3}$ $L/\Delta t$. The dashed line corresponds to a flow rate of $\left(1.20\pm 0.01\right)\times 10^5$  particles/$\Delta t$, which corresponds to a wall shear rate of $\left(8.3\pm 1.4\right)\times 10^{-3}$ $\Delta t^{-1}$ and slip velocity of $\left(7.6\pm 1.2\right)$ $L/\Delta t$.  The minimum number of samples used to calculate the correlation function was $40000$ for the measurement at $1600$ $\Delta t$}
\label{fig:Fig3}
\end{figure}

\begin{figure}[htb]
\includegraphics[width=\columnwidth]{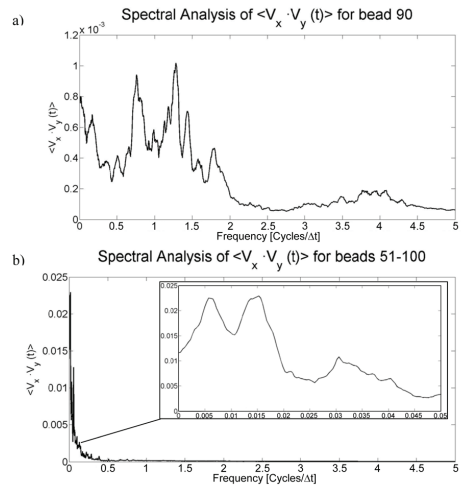}
\caption{Spectral analysis of the cross-correlation function between the $x$ and $y$ velocity components.  The top graph is the function for the $90^{th}$ bead in chain counting from the anchor bead of the chain.  The bottom graph was constructed by averaging the velocity of the last 50 beads of the chain prior to the construction of the cross-correlation function.  Both of these graphs were calculated from the run with a flow rate of $\left(3.26\pm 0.10\right)\times 10^4$  particles/$\Delta$t in \ref{fig:Fig3}.}
\label{fig:Fig4}
\end{figure}

Visualizations of the oligomer trajectory revealed that there appeared to be two time scales of motion for the oligomeric chain; a fast primarily torsional mode of motion along the length of the chain that produces little effect on the radius of gyration and a slower bulk movement of the chain that creates larger fluctuations in the radius of gyration.  The faster mode manifests as finer noise in the data and appears to be periodic, as can be seen in the cross-correlation function of velocity components in the direction normal to the direction of flow computed via the equation 
\begin{eqnarray*}
\left\langle v_xv_y\left( t\right)\right\rangle &=&
\frac{1}{\left(t_s-t\right)\sigma_{v_x}\sigma_{v_y}} \times \\
&&\int^{t_s-t}_0\left(v_x\left(\tau\right)-\overline{v_x}\right)
\left(v_y\left(t+\tau\right)-\overline{v_y}\right) d\tau,
\end{eqnarray*}
where $t_s$ is the length of the simulation, $\overline{v_x}$ and $\overline{v_y}$ are the time averaged values of velocity components $v_x$ and $v_y$, and $\sigma_{v_x}$ and $\sigma_{v_y}$ are the standard deviations of $v_x$  and $v_y$.  Periodic oscillations within the $x-y$ plane will be evident as a sinusoidal wave within this function.  Figure \ref{fig:Fig3} illustrates these data as a function of time for two data sets with a low flow rate and a high flow rate.  The oscillations do not seem to be dependent on flow rate, nor are they consistent across simulations.  They appear to be composed of rotations in the frequency range of $1.75\times 10^{-2}$ to $3.25\times 10^{-2}$ cycles/$\Delta t$.  A window-smoothed spectral analysis of the cross-correlation function of the transverse components of the oligomer velocity is shown in Fig.~\ref{fig:Fig4}a.  The spectral analysis of this correlation function has a large amount of noise, though the bulk of the function can be found in three frequency ranges: the $0$  to $10^{-1}$ cycles/$\Delta t$ range (typically around $1.75 \times 10^{-2}$ to $3.25 \times 10^{-2}$ cycles/$\Delta t$ or twice that value),  a wide peak or cluster of peaks around $1.15 $ cycles/$\Delta t$ and another peak around $4$ cycles/$\Delta t$.  These higher frequencies correspond to the normal modes of the oligomeric chain in the absence of fluid interactions.  They also vanish when the velocity of a segment of chain is used instead of a single fluid particle (Fig.~\ref{fig:Fig4}b).  Hence these modes do not seem to contribute to this torsional reptation.

The slower mode is aperiodic and uncorrelated, and is likely created by local fluctuations in the fluid velocity.  As the shear rate increases, the motion is restricted to the end of the chain.  The likely cause of this restriction is the increased configurational freedom of the untethered end of the oligomeric chain relative to the tethered end.  As noted below, the deviations in the flow field created by the oligomeric chain persist to the end of the chain, so it is possible that this effect is assisted by the shielding effect of the oligomer wake.

\begin{figure}[htb]
\includegraphics[width=0.6\textwidth]{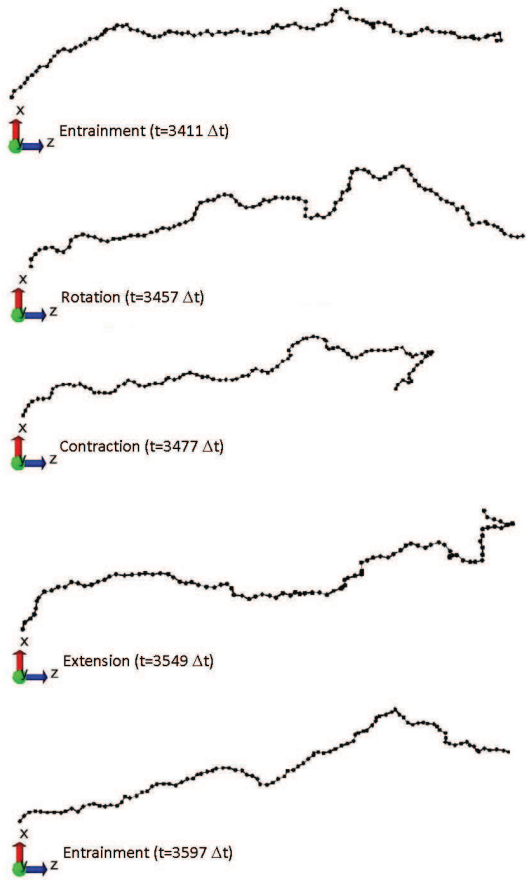}
\caption{Illustration of cyclical dynamics in a simulation with a flow rate of   $\left(1.14\pm 0.01\right)\times 10^5$ particles/$\Delta$t, which corresponds to a wall shear rate of $\left(6.7\pm 1.5\right)\times 10^{-3}$ $\Delta t^{-1}$ and slip velocity of $\left(8.2 \pm 1.7\right)\times 10^{-3}$ $L/\Delta t$.  The cyclical dynamics seem to be confined to the latter quarter of the chain. At $t=3411$  $\Delta t$ the oligomer is extended into the flow field.  At $t=3457$  $\Delta t$ the end of the chain has drifted towards the wall of the channel.  At $t=3477$  $\Delta t$ the end of the chain has retracted.  At $t=3539$ $\Delta t$ the end of the chain has drifted back into the flow field.  At $t=3597$ $\Delta t$ the flow field has extended the chain again.}
\label{fig:Fig5}
\end{figure}

In the introduction, it was noted that cyclical dynamics are often observed for simulations of tethered polymer chains in shear flow.  These were observed in simulations with a flow rate greater than $\left(6.0\pm 0.2 \right)\times 10^4$  particles/$\Delta$t, although the cyclical motion only involves the latter quarter of the chain, instead of the entire chain.  These dynamics are illustrated in Fig.~\ref{fig:Fig5}.  It is worth noting that these dynamics persist in the presence of slip flow, which should hamper the retraction of the oligomeric chain towards the anchor point of the chain.  However, the slip flow is likely the reason why the effect is limited to the end of the chain.  

\begin{figure}[htb]
\includegraphics[width=\columnwidth]{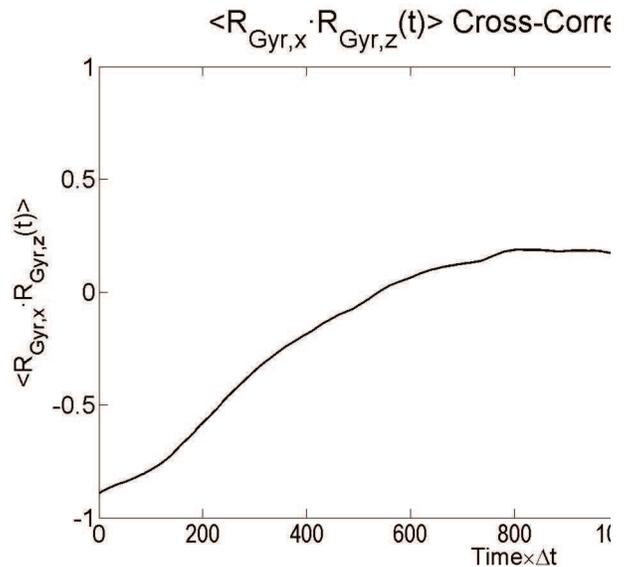}
\caption{The Cross-Correlation function between the x and z components of the Radius of Gyration for the simulation conditions in Fig.~\ref{fig:Fig5}.}
\label{fig:Fig6}
\end{figure}

\begin{figure}[htb]
\includegraphics[width=\columnwidth]{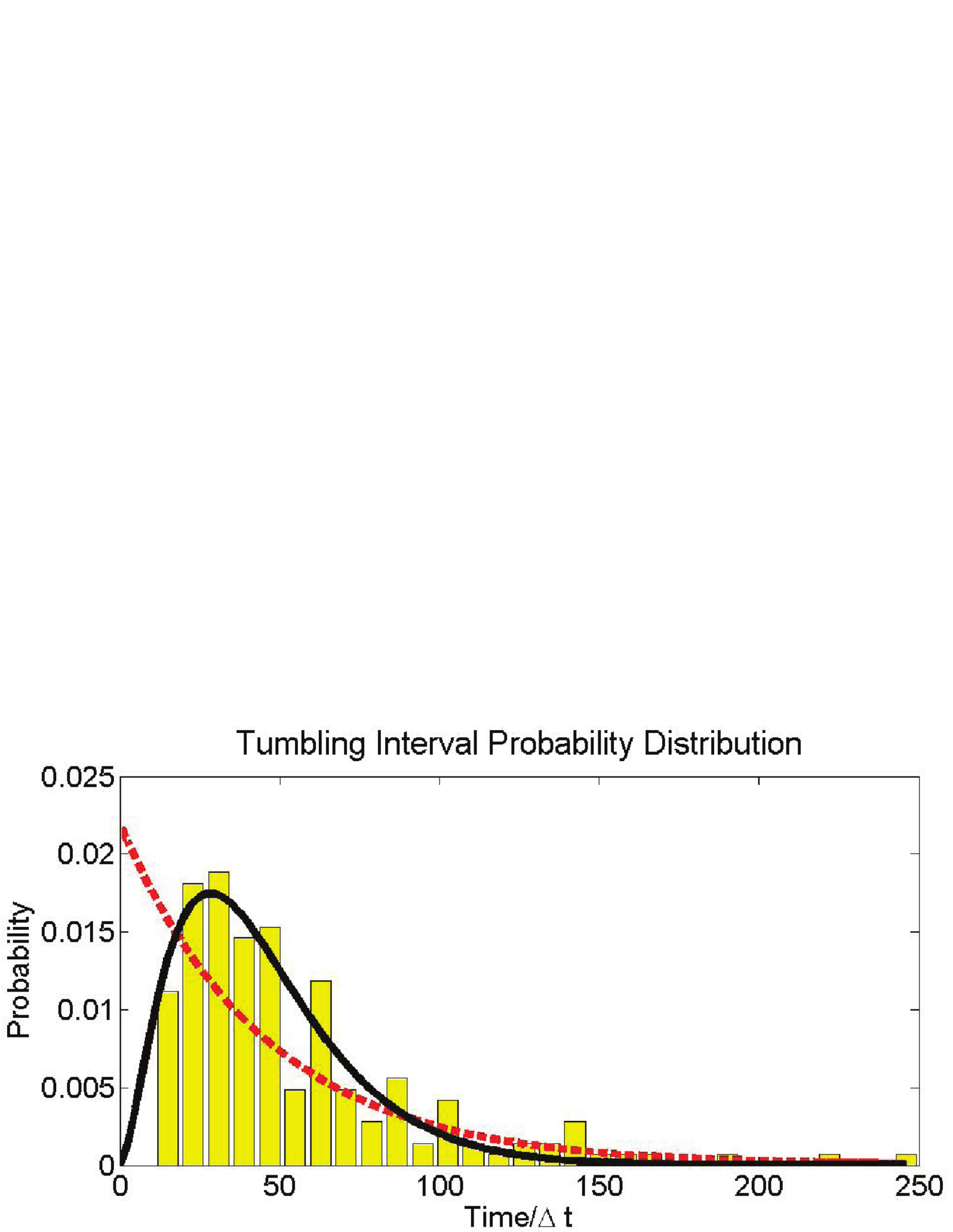}
\caption{The distribution of time intervals between cyclical dynamics events for a set of simulations with flow rate of $\left(1.201\pm 0.007\right)\times 10^5$ particles/$\Delta$t, which corresponds to a wall shear rate of $\left(7.6\pm 1.1\right)\times 10^{-3}$ $\Delta t^{-1}$ and slip velocity of $\left(8.3 \pm 1.4\right)\times 10^{-3}$ $L/\Delta t$.  The distribution, shown as a bar graph has been fit to an exponential distribution (dashed line) and a gamma distribution (solid line).  The exponential distribution has a parameter of $\left(2.2\pm 0.2\right)\times 10^{-2}$ $\Delta t^{-1}$.  The gamma distribution has a shape parameter of $2.7\pm 0.7$ and a scale parameter of $\left(6.0\pm 2.0\right)\times 10^{-2}$ $\Delta t^{-1}$.}   
\label{fig:Fig7}
\end{figure}

It should be noted that these dynamics do not appear to be periodic.  As was noted by Delgado et al \cite{Delgado2006}, if the dynamics were cyclical this should be apparent in the cross-correlation function of the $x$ and $z$ components of the radius of gyration.  However, this cross-correlation function is unremarkable, Fig.~\ref{fig:Fig6}.   It has been suggested \cite{Zhang2009}, that the tumbling of the end of the chain may be a Poisson process.  If that is the case, then the distribution of the time intervals between tumbling events should follow an exponential distribution, Fig.~\ref{fig:Fig7}. Due to thermal noise, the tumbling events are difficult to identify except by visual inspection of animations of the configurations, as shown in Fig.~\ref{fig:Fig5}, where the coordinated motion of the oligomer becomes apparent.  The configurations associated with the \textit{contraction} phase of the cyclical dynamics are visually distinct and short lived during the simulations, hence the time intervals between  \textit{contractions} were used to define the intervals between tumbling events.  At lower flow rates the tumbling events are less distinct, hence this process was restricted to a few high flow rate simulations.   The resulting distribution is modeled better by a gamma distribution with a shape parameter of $2.7\pm 0.7$ and a scale parameter of $\left(6.0\pm 2.0\right)\times 10^{-2}$ $\Delta t^{-1}$ than an exponential distribution.  As the tumbling of the end of the chain requires a certain amount of time and a subsequent tumbling event cannot begin until the previous event has been completed, there are memory effects in the distribution of tumbling events, and the gamma distribution is the correct form to observe.  

A relationship between the relaxation time of the oligomer and the scale parameter of the gamma distribution could not be determined.  The relaxation time of the oligomer, as calculated by fitting an exponential to the z-z correlation function for the terminal bead of the chain \cite{Zhang2009}, was very noisy and could best be empirically modeled as $\epsilon\exp\left(-\dot{\gamma}/\zeta\right)$ where
$\epsilon=\left(4.0\pm 1.2\right)\times 10^{2}$ $\Delta t$, $\zeta=\left(1.3\pm 0.9\right)\times 10^{-2}$ $\Delta t^{-1}$ and $\dot{\gamma}$ is the wall shear rate of the simulation.  Due to the quality of the relaxation time data, a functional dependence between the relaxation time and the exponential parameter could not be determined.  However, if the tumbling mechanism for the end of the chain is similar to the tumbling mechanism for the tumbling times of an untethered polymer in shear flow, then the distribution can compared to the Rouse model of Das and Sabhapandit \cite{Das2008} which models the density of tumbling times,  $p_t(t)$, as $p_t\left(t\right)=t_o/\alpha\exp\left(-\alpha t/t_o\right)$ where $t_o$ is the longest relaxation time of the oligomer.  For the purposes of comparison, the distribution was modeled by an exponential with a parameter value of $\left(2.2\pm 0.2\right)\times 10^{-2}$ $\Delta t^{-1}$.  Using this model, $\alpha\approx 5$ which is significantly larger than the value of Das and Sabhapandit $\alpha=0.325$.  The value predicted by Das and Sabhapandit is a lower bound on the value of $\alpha$.  Semi-flexible polymers are expected to have a higher value of $\alpha$\cite{Celani2005}. The cyclical dynamics appear to occur more rapidly than the tumbling of an oligomer in free solution because the dynamics are restricted to the end of the oligomer instead of the entire oligomer, which requires less motion of the chain for a cycle to occur, and because the oligomer is tethered to the wall of the channel, which increases the tension in the chain.

The relative contribution of the slip flow relative to the shear flow near the wall can be seen by examining the fluid velocity near the wall.  By Taylor series expansion, the fluid velocity near the $x$-normal wall that intercepts the x-axis at $x=-10$ $L$ is approximately 
\begin{equation*}
v_z\left(x\right) \approx v_{slip}+\left( x+10\right)\frac{dv_z}{dx}\bigg\vert_{\text{wall}},
\end{equation*}
 where $v_{slip}$ is the slip velocity and $\frac{dv_z}{dx}\vert_{wall}$ is the shear rate of the fluid at the wall.  From Maxwell's first order model for slip flow\cite{Maxwell1879,Arkillic1997} $v_{slip}=\alpha\lambda \frac{dv_z}{dx}\vert_{wall}$, where $\alpha=0.96\pm 0.04$ (std. error) is the streamwise momentum accommodation and $\lambda=1.11\pm 0.16$ $L$ (std. error) is the mean free path length.  Thus, $v_z\left(x\right) \approx \left( 1.07+x+10\right)\frac{dv_z}{dx}\vert_{wall}$.  As the oligomer is localized between $x=-8$ and $x=-9$ $L$, the slip velocity accounts for approximately $33-50\%$ of the flow field in which the oligomer is entrained.  Although the slip flow is not the dominant portion of the flow field, it is a significant contribution.  A streamwise momentum accommodation of $1$ is a commonly observed value\cite{Arkillic1997} in channels and this slip contribution is reasonable in the vicinity of the polymer.  It is thus surprising that the contribution of slip flow is often neglected in the study of flow on tethered polymers, when its presence should be expected in any simulation where the Knudsen number is greater than $0.001$.

\begin{figure}
\includegraphics[width=0.8\columnwidth]{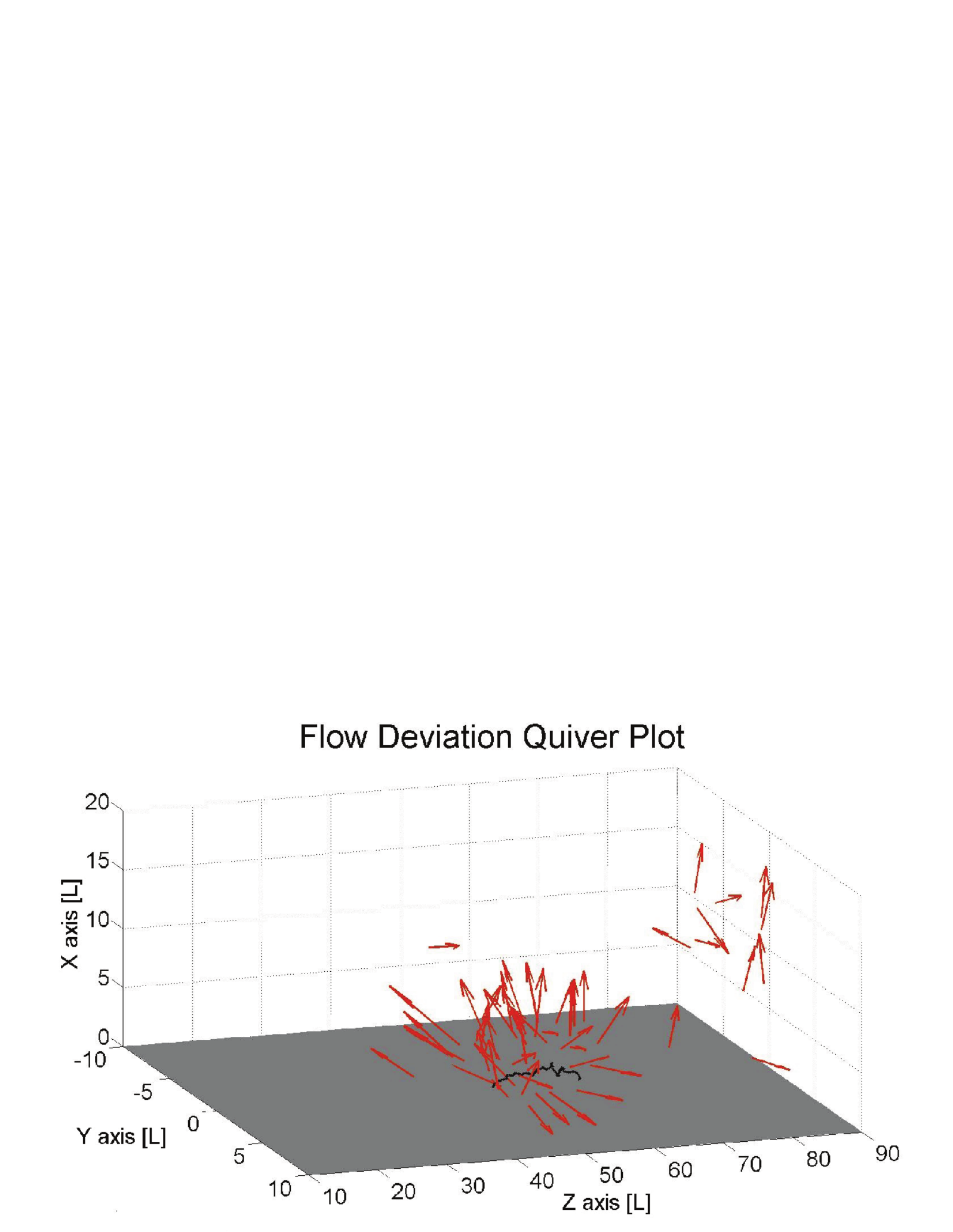}
\caption{A quiver plot of the deviations in the flow field due to the addition of the oligomeric chain for the HF simulation in Fig.~\ref{fig:Fig4}. The data have been time averaged for $400$ $\Delta t$ and any deviations with a magnitude less than $4\times 10^{-3}$ $L/\Delta t$ are not shown in the plot.  The deviations at the end of the channel are due to the error in the fit to the flow field which tend to be larger at the end of the channel than at the beginning.}
\label{fig:Fig8}
\end{figure}

Fig.~\ref{fig:Fig8} shows a quiver plot of the deviations from the normal flow field created by the oligomeric chain.  The plot was created by comparing the flow profile to a fit of the stationary flow profile in the absence of the oligomeric chain to the following function:
\begin{eqnarray*}
v_z\left(x,y,z\right)&=&\sum^6_{i=0}\alpha_iz^i
\bigg(1+\sum^3_{j=1}\beta_j\left(x^{2j}+y^{2j}\right)\\
&&+
\sum^3_{k=1}\sum^3_{l=k}\gamma_{kl}\left(x^{2k}y^{2l}+x^{2l}y^{2k}\right) \bigg)
,
\end{eqnarray*}
where $\alpha$, $\beta$  and $\gamma$  are adjustable parameters.  The deviations were calculated by subtracting the fitted curve from simulation data with the oligomer that had been time averaged over $400$ $\Delta t$.  These deviations were further filtered by removing any deviations that had a magnitude of less than $4\times 10^{-3}$ $L/\Delta t$, which excluded most of the standard noise from the data.  However, the filtering also restricted the analysis to strong flow fields, where the threshold value of the filter would not exclude the data.  The time averaging of the data was necessary due to the thermal noise (since at any instant the thermal fluctuations are on the order of $1.2456\times 10^{-1}$ $L/\Delta t$).  The velocity data were calculated based on particle velocities in the MPCD cells.  As these cells are $L^3$, this defines the maximum resolution of the velocity data and the hydrodynamic interactions within the system.   Due to the the choice to simulate pressure-driven flow of a gas-like solvent as opposed to a liquid-like solvent, the hydrodynamic interactions will be relatively weak.  Since the oligomer beads are lighter than the fluid particles and the thermal fluctuations are relatively large, it would require the coordinated motion of several beads to produce detectable hydrodynamics within the system.  As such, a $L^3$ resolution is quite reasonable.  The deviations show a modification of the flow field around the oligomer that begins before the flow contacts the chain.  The flow deviates in both the $x$ and $y$ directions, like a semi-circular dome before the oligomer.  There is also a decrease in the $z$  component of the velocity.  The deviations persist for the entire length of the oligomeric chain, including an acceleration of the fluid after it has passed the chain, possibly indicating a low pressure area at the end of the chain.  For a real fluid, the disruption to flow created by a single oligomeric chain would be larger than shown in these simulations.  While the MPCD collision scheme allows for momentum transfer between particles, it does not include excluded volume interaction, which would be significant as the fluid is compacted against the oligomeric chain and has to divert.  

\section{Poor Solvent Conditions}

\begin{figure}
\includegraphics[width=\columnwidth]{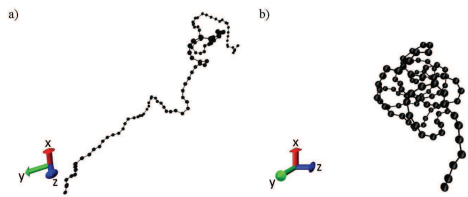}
\caption{Equilibrium oligomer configurations, a (left) ideal solvent b (right) poor solvent.}
\label{fig:Fig9}
\end{figure}

\begin{figure}
\includegraphics[width=0.9\columnwidth]{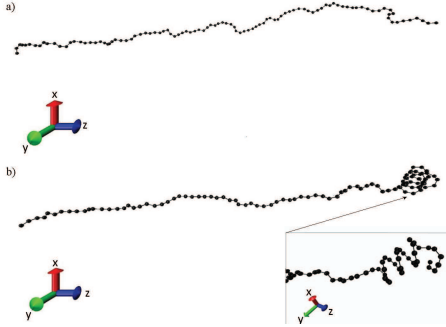}
\caption{Oligomer configurations subject to flow under different solvent conditions.  Top: ideal solvent conditions simulation with a flow rate of $\left(1.20\pm 0.01\right)\times 10^5$ particles/$\Delta t$, which corresponds to a wall shear rate of $\left(8.6\pm 1.8\right)\times 10^{-3}$ $\Delta t^{-1}$ and slip velocity of $9.1\pm 1.4$ $L/\Delta t$.  Bottom: poor solvent simulation with a flow rate of $\left(1.56\pm 0.01\right)\times 10^5$  particles/$\Delta t$, which corresponds to a wall shear rate of $\left(10.1\pm 1.4\right)\times 10^{-3}$ $\Delta t^{-1}$ and slip velocity of $\left(11.8\pm 1.6\right)\times 10^{-3}$ $L/\Delta t$.}
\label{fig:Fig10}
\end{figure}

\begin{figure}[htb]
\includegraphics[width=\columnwidth]{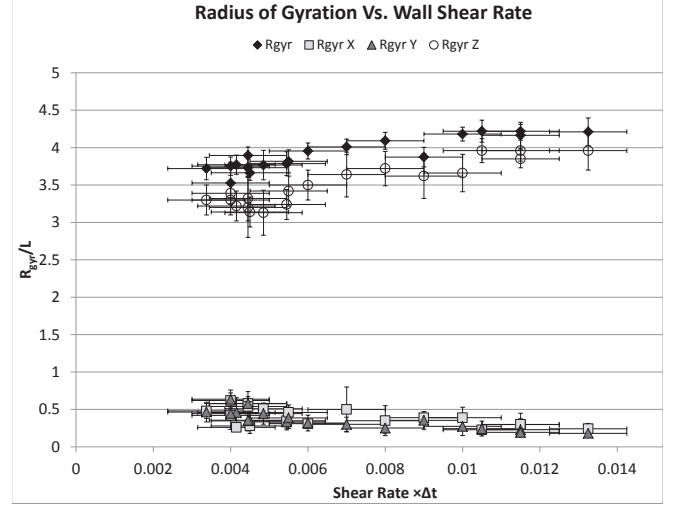}
\caption{$R_{gyr}$ and its three components vs. Wall Shear Rate at the point of adhesion (-10 L, 0 L, 50 L).}
\label{fig:Fig11}
\end{figure}

The differences in the dynamical behavior of a tethered oligomer are quite stark between poor solvent conditions and ideal solvent conditions.  Though an ideal solvent is often considered a poor solvent, it is the threshold between a poor solvent and a good solvent and is considered in essence to be a neutral solvent.  The equilibrium configurations between the poor solvent and ideal solvent are contrasted in Fig.~\ref{fig:Fig9}.   The key difference in the behavior of the oligomer between the poor solvent and ideal solvent conditions is that the oligomeric chain tends to adopt an extended helical conformation or collapsed globule configuration under the poor solvent conditions, as opposed to a random coil that extends into the direction of flow, Fig.~\ref{fig:Fig10}. The radius of gyration of the oligomer and its three components are plotted versus the wall shear rate at the point of adhesion in Fig.~\ref{fig:Fig11}, and exhibit a similar response to the ideal solvent conditions. 

At low flow rates the conformation of the oligomer undergoes a transition from the extended initial configuration directly to a collapsed state.  As the flow rate increases, the oligomer typically adopts a metastable helix conformation extending into the direction of flow, which collapses to the globule state over time ($4000-6000$ $\Delta t$ of simulation in some cases).  At higher flow rates, the oligomeric chain adopts the helix as a stable configuration for the lifetime of the simulations.  The formation of metastable helices as a homopolymer collapses from a stretched initial configuration has been noted in a poor solvent under equilibrium conditions at very low temperatures where the thermal fluctuations are small enough to allow the metastable state to persist \cite{Sabeur2008}, but this is first study in which flow is shown to stabilize this structure.  Furthermore, in the study of Sabeur the poor solvent conditions were simulated via an attractive Lennard-Jones intrapolymer potential that effective allows polymer-polymer interactions to stabilize a compact helical structure, while no explicit attractive intra-polymer interactions exist within the helices presented here.

\begin{figure}
\includegraphics[width=\columnwidth]{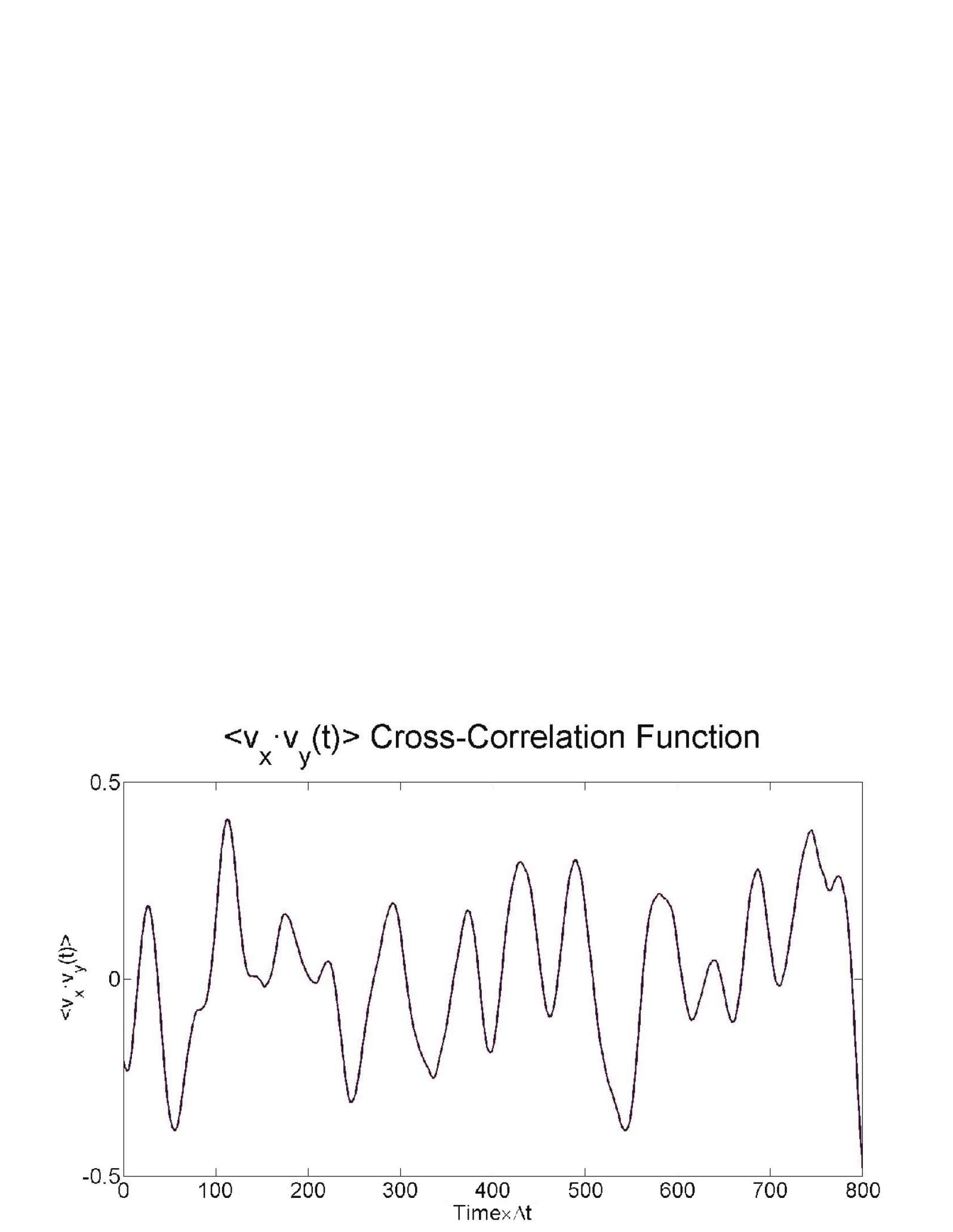}
\caption{Cross-correlation function between the $x$ and $y$ velocity components for the last $50$ beads of the centimer.  The simulation was conducted with a flow rate of $\left(1.56\pm 0.01\right)\times 10^5$ particles/$\Delta t$, which corresponds to a wall shear rate of $\left(10.1\pm 1.4\right)\times 10^{-3}$ $\Delta t^{-1}$ and slip velocity of $\left(11.8\pm 1.6\right)\times 10^{-3}$  $L/\Delta t$.}
\label{fig:Fig12}
\end{figure}

The helix itself exhibits no preference for a left-handed or right-handed configuration.  There appeared to be 9 monomers per turn of the helix, which generally grows from the end of the chain.  Typically there were 3 to 6 turns in the chain, though that number fluctuated over time.  The helix itself forms quite consistently, and a plot of the cross-correlation function of the $x$ and $y$ velocity component of the end of the chain during helix formation show a cyclical pattern consistent with the wrapping of the coil, shown in Fig.~\ref{fig:Fig12}.  The helix formed consistently with a twisting frequency in the $\left(1.9\pm 0.6\right)\times 10^{-2}$ cycles/$\Delta t$ range for all the simulations where the oligomer adopted this conformation.

There were no cyclical dynamics observed for the oligomeric chain in these simulations.  The formation of a helix is surprising given the absence of an attractive potential between oligomer beads.  A visualization of the solvent particles in the vicinity of the oligomer did not indicate any increase in the solvent concentration within the helix or any solvent organization which could have been acting as a bridge to stabilize the helix.

\begin{figure}
\includegraphics[width=0.8\columnwidth]{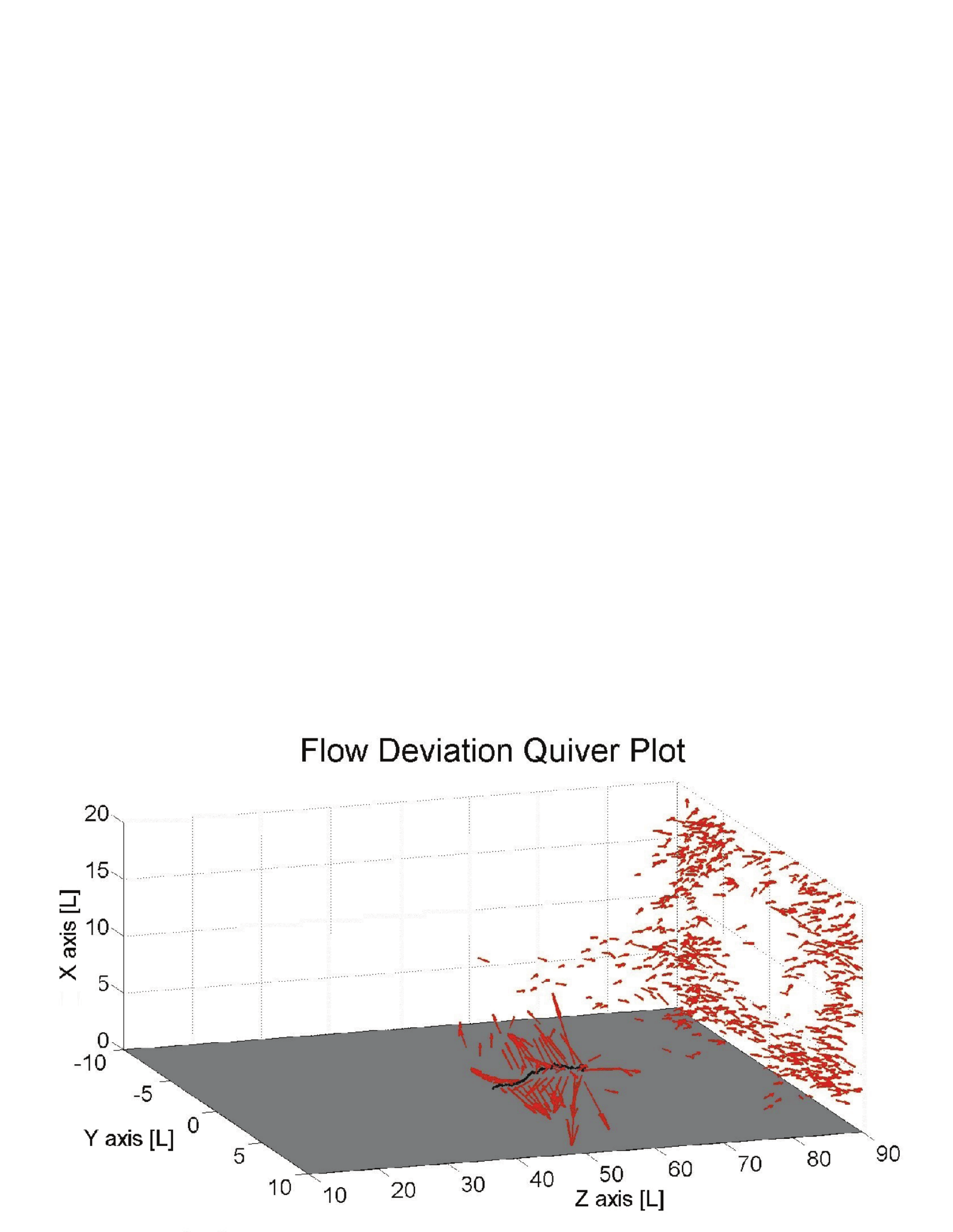}
\caption{A quiver plot of the deviations in the flow field due to the addition of the oligomer chain for the simulation in Fig.~\ref{fig:Fig12}. The data has been time averaged for $400$ $\Delta t$ and any deviations with a magnitude less than $4\times 10^{-3}$ $L/\Delta t$ are not shown in the plot.  The deviations at the end of the channel are due to the error in the fit to the flow field which tend to be larger at the end of the channel than at the beginning}
\label{fig:Fig13}
\end{figure}

The deviations in the flow field created by the oligomer are consistent with the deviations under good solvent conditions, Fig.~\ref{fig:Fig13}.  However, it appears that the largest deviations occur at the end of the chain instead of the beginning, this is likely due to the helix conformation that the chain adopts.  It presents a larger cross-section against the flow than the chain at the adhesion point and hence a larger disruption to the flow profile.

The helical structures are most likely a result of the solvent-induced collapse of the chain from its extended initial conformation to a globule being countered by drag exerted by the flow field on the oligomer.  This hypothesis is supported by the following observations: The helix collapsed to the globule state over time and did not form if a collapsed initial configuration was used instead of an extended configuration or in the absence of the flow; There is no evidence of the solvent stabilizing helical conformations through cross-linking interactions.  The chain begins in an extended configuration and attempts to collapse to the globule state.  However, the flow field orients the chain and exerts a force that keeps the chain extended.  As the flow field adapts itself to the oligomer, the end of the chain is caught in the wake the chain generates and it can begin to collapse due to repulsive interactions with the solvent.  However, it can only collapse within the wake of the chain, which is a narrow cone along the axis of the extended chain.  The torsional reptation of the chain, coupled with the linear path along which the chain can collapse, causes a local collapse of the end of the chain into a helix.  In essence, the helix is the result of a solvent induced polymer collapse with dynamic confinement.  

\section{Conclusions}

The effect of compressible gas flow on an oligomer tethered to the surface of a nanoscopic channel was examined using a hybrid MD-MPCD algorithm.  Channel flow was simulated by establishing a pressure gradient at the inlet and outlet as opposed to a more artificial method that would not replicate both the elongation and shear components of the flow field.  While this model ignored the effects of explicit fluid-fluid interactions, it did incorporate explicit fluid-oligomer interactions and, through the MPCD rotation step, momentum transfer between fluid particles.   As such, it was possible to not only look at the effects of the flow field on the oligomer, but also the effect of the oligomeric chain on the flow field.  Though the wake was limited by the gaseous nature of the solvent and the lack of excluded volume interactions, it still gave an indication of the effect of the oligomer.  The simulations were conducted under two solvent conditions, an ideal solvent condition and a poor solvent condition.  The response of the fluid to the oligomer was consistent under the two solvent conditions but the response of the oligomer to the fluid was quite different.

The oligomer behaved in a manner consistent with previous observations under ideal solvent conditions.  The oligomer extended into the direction of flow to a value roughly equal to 80 \% of its extension length.  Aperiodic cyclical dynamics were also observed at the higher flow rates as the oligomer extended into the flow field.  The distribution of the intervals between cycles yielded a gamma distribution.  As the flow has an elongational component and a slip component, it is surprising that these dynamics persist as it is variation of the shear force as a function of distance from the wall that is believed to give rise to these dynamics.  It is likely that the variational component of the flow is large enough relative to the constant component of flow to allow these dynamics to persist.

The flow field prompted a transition between a collapsed globule configuration for the oligomer and an extended helix state under poor solvent conditions.  Due to the initial configuration that extended the oligomer into the flow field, there were some realizations of the dynamics where the oligomer initially transitioned into the helix state which then collapsed into the globule state, though at higher flow rates the oligomer remained extended.

These results, in particular under poor solvent conditions, indicate that the effect of flow rate on the conformation of macromers could be a design consideration for lab-on-a-chip devices.  In particular, flow induced denaturation of proteins or surface active molecules could limit flow rates in some of these devices.  However, there are also potential advantages to flow induced conformational changes.  Polymerization of monomers under high shear solvent conditions could result in the creation of highly crystalline aligned polymers; it may even be possible through flow modulation to create varying layers of crystalline and amorphous polymer in the same material during the same polymerization.

The hybrid MPCD-MD algorithm used in this article to simulate an oligomer in a nanoscopic channel is only one potential application for this simulation technique.  It would be inefficient to use this method to simulate dense polymer networks where the MD subsystem would occupy the majority of the simulation space, but the technique is practicable for the simulation of channels that are sparsely populated with tethered polymers, or to examine alternate geometries (e.g. a polymer chain on the inside of a right angle bend in a channel). 

\section*{Acknowledgments}
This work was supported by a grant from the 
Natural Sciences and Engineering
Research Council of Canada.

\end{document}